\documentclass[aps,physrev,preprint,groupedaddress]{revtex4-2}

\usepackage{enumerate}
\begin{document}


\title{Comment on ``Consequences of the single-pair measurement of the Bell parameter''}


\author{Justo Pastor Lambare}
\email[]{jupalam@gmail.com}
\affiliation{Universidad Nacional de Asuncion-FACEN, Ruta Mcal. J. F. Estigarribia, Km 11 Campus de la UNA, San Lorenzo-Paraguay}


\date{\today}

\begin{abstract}
In a recent article [Phys. Rev. A 111, 022204 (2025)], Genovese and Piacentini analyzed recent experiments measuring what they call ``the entire Bell-CHSH parameter.''
They claimed those experiments may have implications for interpreting loophole-free tests of the Bell-CHSH inequality.
We explain that the Bell-CHSH inequality is not based on the entire Bell parameter, so these experiments are unrelated to their empirical tests and cannot close any eventual loopholes that might still persist.
We point out that the physical meaning of these new experiments measuring the entire Bell parameter could be interpreted differently.
\end{abstract}


\maketitle


In reference \cite{pG&P25} Genovese and Piacentini (G\&P)  claim that recent weak-interaction-based experiments \cite{pVir24} have consequences for the Bell-CHSH inequality and challenge some statements regarding its interpretation.

Their claim is based on the measurement of the entire Bell-CHSH parameter.
According to G\&P the Bell-CHSH inequality is expressed as $|\mathcal{B}|\leq2$, where $\mathcal{B}$ is the Bell-CHSH parameter,
\begin{equation}\label{eq:G&PBP}
\mathcal{B}=\langle A_1B_1 + A_1B_2 + A_2B_1 - A_2B_2\rangle
\end{equation}
The four values $A_iB_k$ are supposed to be measured only on one single pair of entangled particles as explained in \cite{pG&P25}:
\begin{quote}
\textit{This implies that each copy of the bipartite entangled state exploited actually contributes to estimate just one of the four terms of the Bell-CHSH parameter $\mathcal{B}$ in Eq. (\ref{eq:G&PBP}), since the quantification of the entire parameter is forbidden by the impossibility of measuring, at once, all the observables needed for such a task.}
\end{quote}
According to the above description, the Bell-CHSH inequality would be empirically unrealizable.
On the other hand, the correct version of the Bell-CHSH inequality is expressed as $|\hat{S}|\leq2$, where
\begin{equation}\label{eq:hS}
\hat{S}=\langle A_1B_1\rangle +\langle A_1B_2\rangle +\langle A_2B_1\rangle -\langle A_2B_2\rangle
\end{equation}
The physical meaning of $\hat{S}$ is:
\begin{enumerate}[(A)]
\item $\hat{S}$ is obtained after a long series of experiments allowing sufficient data to evaluate $\langle A_iB_k\rangle$ for each possible combination of settings through the individual values  $A_iB_k$ obtained by actually measuring only once the spin of each particle of the singlet state.
\end{enumerate}
Thus, we see that the Bell-CHSH parameter $\mathcal{B}$, Eq. (\ref{eq:G&PBP}), is not related to the Bell-CHSH inequality and is required neither for its experimental implementation nor for assessing its implications.
It is convenient that we have a primary source to clarify that (\ref{eq:G&PBP}) is not related to the Bell-CHSH inequality.
The first criticism of the Bell-CHSH inequality based on an entire Bell parameter kind of interpretation was made in 1972 \cite{pDCB72}.
Under the requirement of an editor, Bell responded to the criticism  \cite{pBel75}:
\begin{quote}
\textit{The objection of de la Pe\~{n}a, Cetto, and Brody is based on a misinterpretation of the demonstration of the theorem......But by no means. We are not at all concerned with sequence of measurements on a given particle, or of pairs of measurements on a given pair of particles. We are concerned with experiments with which for each pair the ``spin'' of each particle is measured once only.}
\end{quote}
Bell's above laconic explanation to De La Pe\~{n}a et al. was not enough to prevent it from becoming a recursive misinterpretation, which became an endemic problem for correctly assessing the Bell inequality implications.
A detailed discussion of this problem is presented in Ref. \onlinecite{pLam21b} and also a response to a similar more recent claim by Cetto et al. \cite{pCet20} in Ref. \onlinecite{pLam22a}.

Finally, we point out that G\&P entire Bell-CHSH parameter can be associated with a different inequality proposed by Stapp in 1971 \cite{pSta71}.
Inspired by the Bell-CHSH inequality, but contrary to it, the Stapp inequality is based on counterfactual reasoning and does not assume hidden variables.
Stapp used his inequality to prove a theorem on quantum nonlocality, while the Bell-CHSH inequality is used to prove a no-local hidden variables theorem.
Stapp interprets the expression,
\begin{equation}\label{eq:stapi}
A_1B_1+A_1B_2+A_2B_1-A_2B_2
\end{equation}
according to the following description \cite{pSta71}:
\begin{quote}
\textit{Of these eight numbers only two can be compared directly to experiment. The other six correspond to the three alternative experiments that could have been performed but were not.}
\end{quote}
Thus, G\&P Bell-CHSH parameter given by (\ref{eq:G&PBP}) is the same as Stapp's description of (\ref{eq:stapi}).
Table \ref{tabla:t1} shows the differences between Bell-CHSH and Stapp's inequalities.
The conceptual differences between Bell and Stapp's inequality are further discussed in an appendix to Ref. \onlinecite{pLam22b}.
\begin{table}[t]
\caption{Bell-CHSH vs. Stapp Inequality\label{tabla:t1}}
\begin{center}
\begin{tabular}{|c|c|c|c|}
\hline
Inequality & Hidden     & Quantum     & Counterfactual \\
  Type     & Variables  & Nonlocality &    Reasoning   \\
\hline
B-CHSH     &   Yes      & No          &   No           \\
\hline
Stapp     &    No       & Yes         &   Yes          \\
\hline
\end{tabular}
\end{center}
\end{table}
\section*{Conclusion}
The weak-interaction-based experiments approach to the Bell-CHSH inequality cannot improve on the loophole-free experiments and the certainty obtained from them about the untenability of local realism unless we accept superdeterminism.

However, the usefulness of these new experiments based on weak measurements applied to Bell-type inequalities can not be discarded, particularly for quantum foundations; they could have consequences for Stapp's approach to quantum nonlocality, perhaps as a possible actual implementation of counterfactual reasoning.

\bibliography{zBellbibfile}

\end{document}